\documentclass[12pt,a4paper]{article}
\usepackage{amsmath}

\begin{document}

\title{The non-chiral fusion rules \\
in rational conformal field theories}
\author{A. Rida  \\
%EndAName
{\small Laboratoire de PhysiqueTh\'{e}orique, Institut de Physique,}\\
{\small Universit\'{e} des Sciences et de la Technologie Houari Boumedienne,}%
\\
{\small BP 32 El-Alia, Alger 16111, Algeria} \and T. Sami\thanks{%
e-mail: sami@subatech.in2p3.fr } \\
%EndAName
{\small Laboratoire SUBATECH, Universit\'e de Nantes, Ecole des Mines de Nantes, }\\
{\small 4 Rue Alfred Kastler, La Chantrerie 44070 Nantes, France}}
\date{}
\maketitle

\begin{abstract}
We introduce a general method in order to construct the non-chiral fusion
rules which determine the operator content of the operator product algebra
for rational conformal field theories. We are particularly interested in the
models of the complementary $D-$like solutions of the modular invariant
partition functions with cyclic center $Z_{N}$. We find that the non-chiral
fusion rules have a $Z_{N}-$grading structure.
\end{abstract}

One of the most important requirements in the construction of
two-dimensional conformal field theories (CFT), the bootstrap requirement,
is the existence of a closed associative operator product algebra (OPA)
among all fields \cite{BPZ}. The information about the OPA structure is
collected in the structure constants, the computation of which is fundamental
since it permits in principle the determination of all correlation
functions. But unfortunately, in practice, the structure of the OPA seems to
be very complicated and so is, indeed, the computation of the structure
constants.

The structure of the OPA is expressed by the so-called \textit{fusion rules} 
\cite{Fusion}, which describe how the chiral-sectors `vertices',
representing the holomorphic $\left( z\right) $ and antiholomorphic $\left( 
\overline{z}\right) $ dependencies of the primary fields, combine in the OPA 
\cite{Fuchs}. Although these fusion rules contain a large amount of
information, this description seems to be
incomplete. Indeed, for the maximally extended chiral algebras, i.e., models
with no pure holomorphic primary fields, the chiral fusion rules describe
completely the OPA structure since the two structures are isomorphic \cite
{Moore}. But for the non-maximally extended algebras, the relationship
between the two structures is still not clear and in these special cases,
the chiral fusion rules describe in a `chiral' half-manner only the full OPA
structure \cite{Moore}.

The purpose of this letter is to propose a construction that provides a
complete description of the OPA structure using the notion of the \textit{%
non-chiral} fusion rules, first introduced in \cite{John, Amar} for the $D$
series of the minimal $\left( c<1\right) $ models. The non-chiral fusion
rules, in contrast to the chiral ones, determine the operator
content of the operator product algebra, i.e., they determine which fields
are present in the product (fusion) of two primary fields.

The first step is of course the chiral fusion rules: a non-chiral fusion
coupling is vanishing if and only if the corresponding chiral fusion
coupling in the two sectors vanishes. This is called the \textit{naturality}
statement \cite{Moore}. The second step consists in combining properly the
fusion rules in the two sectors by considering the closure of the OPA so
that only combinations permitted by the operator content are considered. We
note that the operator content, in relation with the modular solutions of
the partition function, has been completely determined for a particular
class of rational conformal theories (RCFT) known as ``simple current
solutions'' \cite{Simple1, Simple2}. We will henceforth specialize to this
particular type of models. The third step is to show how the "simple-current
symmetry", a symmetry of the fusion rules, manifests itself in the
construction of the non-chiral fusion rules. In this letter, we restrict
ourselves to the case where the current symmetry forms a cyclic group $Z_{N}$%
.

Let us now start our construction. The existence of a closed associative
operator algebra among all fields is expressed in the following : 
\begin{equation}
\Phi _{\mathbf{I}}\left( z,\overline{z}\right) \cdot \Phi _{\mathbf{J}%
}\left( w,\overline{w}\right) =\sum_{\mathbf{K}}C_{\mathbf{IJK}}\left(
z-w\right) ^{h_{k}-h_{i}-h_{j}}\left( \overline{z}-\overline{w}\right) ^{%
\overline{h}_{k}-\overline{h}_{i}-\overline{h}_{j}}\left[ \Phi _{\mathbf{K}%
}\left( w,\overline{w}\right) +\ldots \right] .  \label{OPA}
\end{equation}
$I\equiv \left( i,\overline{i}\right) $ indicates the different fields with $%
i\left( \overline{i}\right) $ representing the contributions (\textit{%
vertices}) of the holomorphic (antiholomorphic) sectors to the primary
fields. $h_{i}$ and $\overline{h}_{i}$ are the conformal dimensions of $\Phi
_{\mathbf{I}}$ and the ellipsis stands for terms involving descendant
fields. $C_{\mathbf{IJK}}$ are the structure constants of the operator
algebra. We define the non-chiral fusion rules as a formal structure which
determine the operator content of the operator algebra. Thus, the fusion of
the two primaries $\Phi _{\mathbf{I}}$ and $\Phi _{\mathbf{J}}$ produces the
field $\Phi _{\mathbf{K}}$ if and only if the structure constant $C_{\mathbf{%
IJK}}$ in (\ref{OPA}) is non-zero. If we adopt $\left( i,\overline{i}\right) 
$ and $\ast $ as notations for a primary field $\Phi _{\mathbf{I}}$ and the
OPA operation respectively, the non-chiral fusion rules are formally given
by: 
\begin{gather}
\left( i,\overline{i}\right) \ast \left( j,\overline{j}\right) =\sum_{%
\mathbf{K}}\mathcal{N}_{\mathbf{IJ}}^{\mathbf{K}}\,\,\,\left( k,\overline{k}%
\right) ;  \notag \\[0.08in]
\mathcal{N}_{\mathbf{IJ}}^{\mathbf{K}}\neq 0\Leftrightarrow C_{\mathbf{IJK}%
}\neq 0,  \label{non-chiral f-r}
\end{gather}
Eq. (\ref{non-chiral f-r}) is similar to how the chiral fusion rules
(describing an associative structure) in the two sectors are usually
defined. Indeed, we usually associate with each primary field $\Phi _{%
\mathbf{I}}$ a formal object $\left( i\right) $ representing its chiral part
and introduce a formal multiplication $\ast $ by: 
\begin{equation}
\left( i\right) \ast \left( j\right) =\sum_{k}\mathcal{N}_{ij}^{k}\,\left(
k\right) ,
\end{equation}
where $\mathcal{N}_{ij}^{k}$ (chiral fusion rules coefficients) counts the
number of distinct ways in which $\left( k\right) $ occurs in the chiral
fusion of the two fields $\left( i\right) $ and $\left( j\right) $
respectively \cite{Fusion}. If we denote by an arrow $\left( \rightarrow
\right) $ the presence of $\left( k\right) $ in the fusion $\left( i\right)
\ast \left( j\right) $, the naturality statement amounts to: 
\begin{equation}
\left\{ 
\begin{array}{c}
\left( i\right) \ast \left( j\right) \rightarrow \left( k\right) \\ 
\left( \overline{i}\right) \ast \left( \overline{j}\right) \rightarrow
\left( \overline{k}\right)
\end{array}
\right. \Rightarrow C_{\mathbf{IJK}}\neq 0,
\end{equation}
which means that for the non-chiral fusion rules, one must have the
condition: 
\begin{equation}
\mathcal{N}_{ij}^{k}\neq 0,\,\,\,\,\mathcal{N}_{\overline{i}\,\overline{j}}^{%
\overline{k}}\neq 0\Rightarrow \mathcal{N}_{\mathbf{IJ}}^{\mathbf{K}}\neq 0.
\label{naturality}
\end{equation}

As mentioned earlier on, the second step is to consider the closure of the
OPA: we require the compatibility of the non-chiral fusion rules with the
operator content determined by the modular constraints on partition
functions. Denoting the set of the operator content of a given conformal
model by $\mathcal{O}$, this requirement is expressed by: 
\begin{equation}
\mathcal{N}_{\mathbf{IJ}}^{\mathbf{K}}\neq 0\,\ \text{\thinspace
if\thinspace }\,\,\,\mathbf{I,J,K\in \,}\mathcal{O}.
\end{equation}

The third step relies, as we said, on symmetry considerations. Indeed, a
symmetry of the fusion rules provided by the presence of \textit{simple
currents} is quite useful in the construction of a large class of modular
invariant partition functions \cite{Simple1, Simple2}. The construction of
simple currents as solutions to modular invariant partition functions
reproduces the diagonal $\left( A\right) $-like and the non-diagonal $\left(
D\right) $-like series of the $SU\left( 2\right) $ Kac-Moody algebra. In the 
$\left( D\right) $ series case, it was emphasized in \cite{John, Amar,
FuchsZ2} that the OPA is structured \ following a discrete cyclic symmetry $%
Z_{2}$. Here we show how the analogous to this symmetry appears in more
general cases as a consequence of a simple-current structure of the $D$-like
series.

But first, a brief reminder of the structure of a simple-current symmetry. A
simple current $\left( j\right) $ is a primary field (chiral part) for which
the chiral fusion rules with $\left( i\right) $ give one non-vanishing
fusion coefficient $\mathcal{N}_{ji}^{k}$, so that $\left( j\right) \mathbf{%
\ast }\left( i\right) =\left( k\right) $ \cite{Simple1}. Due to the
associativity of the fusion product, the fusion of two simple currents is
again a simple current. Simple\ currents thus form an abelian group under
the fusion product, called the \textit{center of the theory}. Since the
number of primary fields is finite in a rational theory, the number of
simple currents is finite. This implies that simple currents are unipotent,
i.e., there must be an integer $N$ so that $j^{N}=1.$ The smallest integer $%
N $ with this property is called the order of the simple current.
Furthermore, by rationality, there must also be a smallest positive integer $%
N_{i}$ such that $\left( j\right) ^{N_{i}}\ast \left( i\right) =\left(
i\right) $. By associativity, $N_{i}$ must be a divisor of $N$. If a simple
current takes a field into itself, we call this latter a \textit{fixed point}
of that current. We see then that any simple current organizes the primary
fields into orbits: $\left\{ \left( j^{\alpha }\ast i\right) ;\,\,\,\,\alpha
=0\ldots N_{i}\right\} $.

The presence of simple currents in a conformal field theory allows not only
the organization of the fields into orbits but also the expression of a
symmetry through a conservation of a monodromy charge $Q\left( i\right) $.
This latter is defined modulo an integer and expresses the monodromy
property of the (OPA) (\ref{OPA}) of a primary field $\left( i\right) $ with
a simple current $\left( j\right) $: $Q\left( i\right) \equiv
h_{j}+h_{i}-h_{\left( j\ast i\right) }%
%TCIMACRO{
%\TeXButton{pmod}{\bmod%
%}}%
%BeginExpansion
\bmod%
%
%EndExpansion
1.$ This definition imposes the monodromy charge to be additive under the
operator product, i.e., $Q\left( i\ast k\right) =Q\left( i\right) +Q\left(
k\right) $ and it follows that all terms in the OPA of two fields \textit{%
must have the same charge}. As a consequence, the monodromy charge must be
conserved $\left( \sum_{n}Q\left( i_{n}\right) =Q\left( 1\right) =0%
%TCIMACRO{
%\TeXButton{pmod}{\bmod%
%}}%
%BeginExpansion
\bmod%
%
%EndExpansion
1\right) $ in order to have a non-vanishing $n$-point correlation function $%
\left\langle i_{1}i_{2}\ldots i_{n}\right\rangle $. So, if we denote the
charge associated with a simple current $j$ by $Q$, then obviously the
charge associated to $j^{d}$ is equal to $dQ$, as usual modulo $1$. This
implies that simple currents of order $N$ $\left( j^{N}=1\right) $ have a
charge equal to an integer, and therefore $Q\left( j\right) \equiv \frac{r}{N%
}%
%TCIMACRO{
%\TeXButton{pmod}{\bmod%
%}}%
%BeginExpansion
\bmod%
%
%EndExpansion
1$, $r$ defined modulo $N$.

The conservation of the monodromy charge means that the presence of simple
currents in a conformal field theory implies that the center forms an abelian
discrete symmetry group. This important fact is very useful in order to find
solutions to a large class of modular invariant partition functions using an 
\textit{orbifold-like} method with respect to the center. The resulting
modular invariant partition functions are known as simple-current
invariants. A modular invariant partition function $\left( Z=\sum
M_{i,k}\chi _{i}\chi _{k}\right) $ is called a simple current invariant if
all fields paired non-diagonally are related by simple currents: $%
M_{i,k}\neq 0$ if $k=ji$. In the case where the center forms a cyclic center $%
Z_{N}$, i.e., there is only one orbit of simple currents $\left\{
1,j,j^{2},\ldots ,j^{N-1}\right\} $, the non-zero couplings $M_{i,k}$ are
given by \cite{Simple1, Simple2}: 
\begin{equation}
M_{i,j^{n}i}=\frac{N}{N_{i}}\delta ^{1}\left[ Q\left( i\right) +\frac{1}{2}%
nQ\left( j\right) \right] ,  \label{Mpart}
\end{equation}
where $\delta ^{1}$ is equal to one if its argument is an integer, zero
otherwise. These solutions (\ref{Mpart}) can be structured as either an
automorphism-invariant and integer-spin-invariant structures or a
combination of these. In the first structure, the two chiral sectors $\left(
i,j\right) $ are combined via an automorphism of fusion rules $\pi \left(
i\right) =j$. The second structure can always be regarded as a diagonal
invariant of a larger than originally considered chiral algebra \cite{Moore}%
. This is what happens when some current $j^{n}$ has an integer spin so that
the coupling $M_{0,j^{n}}$ is different from zero and hence produces an
holomorphic field corresponding to the extra currents (Noether currents)
that will extend the algebra. For the cyclic center $Z_{N}$ \ with $N$
prime, one has a pure automorphism invariant for $Q\left( j\right) \neq 0$
or integer invariant for $Q\left( j\right) =0$. For the non-prime values of $%
N$, the general form of modular partition function can be a combination of
the automorphism and integer-invariant solutions. In the sequel, we will
exploit the operator-content structure provided by the forms of the
modular-invariant partition functions for $N$ prime in the construction of
the non-chiral fusion rules. The same conclusions will be straightforwardly
applied to non-prime values of $N$.

We now turn to the construction of the non-chiral fusion rules. We start
with the automorphism invariant cases. The fact that the two chiral sectors
of a primary field are related by an automorphism of fusion rules makes the
OPA structure isomorphic to the structure of the fusion rules \cite{Moore}.
To see this, we first consider the operator content obtained from (\ref
{Mpart}) for $Q\left( j\right) \neq 0$. This content can be structured in
sets: 
\begin{equation*}
A_{k}=\left\{ \left( j^{k}i,j^{N-k}i\right) \,/Q\left( i\right) =0\right\} .
\end{equation*}
One can see that the two sectors are related by an automorphism of fusion
rules $\pi \left( i\right) =i^{c}$ where $i^{c}$ is the conjugate field of $%
i $ under the fusion operation. To construct the non-chiral fusion rules,
let us consider two fields $\left( j^{k}i,j^{N-k}i\right) \,$ and $\left(
j^{p}i^{\prime },j^{N-p}i^{\prime }\right) \,$ of $A_{k}$ and $A_{p}$
respectively. Due to charge conservation in the chiral fusion rules, we
have: 
\begin{eqnarray*}
\left( j^{k}i\right) \ast \left( j^{p}i^{\prime }\right) &=&j^{k+p}\left(
i\ast i^{\prime }\right) \rightarrow j^{k+p}k,\,\,Q\left( k\right) =0; \\
\left( j^{N-k}i\right) \ast \left( j^{N-p}i^{\prime }\right) &=&j^{N-\left(
k+p\right) }\left( i\ast i^{\prime }\right) \rightarrow j^{N-\left(
k+p\right) }k,\,\,Q\left( k\right) =0.
\end{eqnarray*}
The compatibility with the operator content implies then: 
\begin{equation}
\left( j^{k}i,j^{N-k}i\right) \ast \left( j^{p}i^{\prime },j^{N-p}i^{\prime
}\right) =\sum_{k}\left( j^{p+k}\left( i\ast i^{\prime }\right)
_{k},j^{N-\left( p+k\right) }\left( i\ast i^{\prime }\right) _{k}\right) ,
\label{Non1}
\end{equation}
where $\left( i\ast i^{\prime }\right) _{k}$ represents all fields $k$ which
are results of a chiral fusion of $i$ and $i^{\prime }$. In a more compact
form, the non-chiral fusion rules are written as: 
\begin{equation}
A_{k}\ast A_{p}=A_{l},\,\,\,l=p+k%
%TCIMACRO{
%\TeXButton{pmod}{\bmod%
%}}%
%BeginExpansion
\bmod%
%
%EndExpansion
N.  \label{Fusion1}
\end{equation}
The non-chiral fusion rules have then the structure of a $Z_{N}$-grading
where $N$ is the order of the center $Z_{N}$. This structure (\ref{Non1}) is
isomorphic to the chiral fusion rules and thus its associativity derives
directly from the associativity of the fusion rules.

It is important to make here two observations related to the $Z_{N}$
structure (\ref{Fusion1}). First, one can see that the set of scalar fields $%
A_{0}$ forms a closed sub-algebra in the OPA. Secondly, the $Z_{N}$-grading
structure of the non-chiral fusion rules (\ref{Fusion1}) points to the
conservation of another charge (other than the monodromy charge) in the
non-chiral fusion rules. Indeed, if we assign to each field in a set $\left(
A_{p}\right) $ a charge $\left( p/N\right) $ which will be called a ``%
\textit{distance charge}'', this charge is conserved.

The above two observations are very important in the construction of the
non-chiral fusion rules in the integer invariant cases where the problem is
less evident. Indeed, if the automorphism structure discussed so far of the
operator content allows us to have an isomorphism between the OPA and the
fusion rules, this is not the situation in the integer invariant cases.
Formally, the difficulties arise principally from the fact that if $Q\left(
j\right) =0$, then the charge assigned to $\left( j^{p}i\right) $ with $%
Q\left( i\right) =0$ is also zero.

We tackle the issue in these cases starting from the modular partition
functions (\ref{Mpart}) which yields the following structure of the operator
content: 
\begin{equation}
A_{k}=\left\{ \left( i,j^{k}i\right) /Q\left( i\right) =0\right\}
,\,\,k=1\ldots N.  \label{cont2}
\end{equation}
Let us now proceed naively with our algorithm as we did in the automorphism
invariant cases and see what happens. From the structure of the operator
content, one can easily see that it is not possible to have a nice structure
as in (\ref{Fusion1}). Indeed, one can see that for each $\left( i\right) $
with $Q\left( i\right) =0$, we can find $p$ and $\left( k\right) $ with $%
Q\left( k\right) =0$ such that $j^{p}k=i.$ As a consequence, in the results
of the products of the `\textit{scalar}' fields $A_{0}\ast A_{0},$ it is
possible to obtain fields from different sets (\ref{cont2}) which are
not `scalars'. A more serious problem comes from the presence of orbits with
length $N_{i}\neq N$ or, in other words, the presence of fixed-point fields $%
\left( j\right) \ast \left( f\right) =\left( f\right) $. From (\ref{Mpart}),
this means ($N_{i}=1$) that one has $N$ copies of the scalar field $\left(
f,f\right) $ in the operator content. In the structure of the operator
content (\ref{cont2}), this multiplicity translates in that a fixed point of 
$\left( j\right) $ is also a fixed point of $\left( j^{k}\right) $, so that
the fields $\left( f,j^{k}f\right) \in A_{k}$ are equal to $\left(
f,f\right) $. In the construction of the non-chiral fusion rules, the
presence of a multiplicity of a field creates a problem since we are not
able to differentiate the behavior of the different copies of this field in
the (OPA).

Following our early work \cite{Amar}, the existence of multiple copies ($N$
copies) of the same primary field $\left\{ \Phi _{p},p=0\ldots N-1\right\} $
implies, by the principle of correspondence fields-states, that the
corresponding lowest-weight state (ground state vector) is degenerate. This
observation suggests the existence of a discrete symmetry $Z_{N}$ of the
ground states such that: 
\begin{equation}
Z_{N}\left( \Phi _{p}\right) =e^{\frac{i2\pi p}{N}}\Phi _{p}.  \label{Copies}
\end{equation}
Thus the symmetry $Z_{N}$ enables one to separate the contributions from the
different copies of a degenerate field: we only need to impose the
consistency of the non-chiral fusion rules with the action of this symmetry.
But we must also determine the action of $Z_{N}$ on the other fields of the
theory. In order to do this, we first use a heuristic argument, namely that
the fields of the same structure (fields belonging to the same set $A_{p}$)
must have the same behavior under $Z_{N}$. Observing that each copy $\Phi
_{p}$ of a degenerate field belongs to a set $A_{p}$, we can write
that: 
\begin{equation}
Z_{N}\left( \Phi \right) =e^{\frac{i2\pi p}{N}}\Phi ,\,\,\Phi \in A_{p}.
\label{Action}
\end{equation}
The action of the discrete symmetry $Z_{N}$ provides each field from $A_{p}$
with a charge equal to the distance charge $\left( p/N\right) $. The
consistency of the non-chiral fusion rules with the $Z_{N}$ action (\ref
{Copies}, \ref{Action}) implies the conservation of this charge under the
fusion operation, so that finally we obtain a $Z_{N}$-grading (like in the
automorphism cases) of these rules: 
\begin{equation}
\left( i,j^{p}i\right) \ast \left( i^{\prime },j^{p^{\prime }}i^{\prime
}\right) =\sum_{k}\left( \left( i\ast i^{\prime }\right)
_{k},\,\,j^{p+p^{\prime }}\left( i\ast i^{\prime }\right) _{k}\right) .
\label{Fusion2}
\end{equation}
In these rules, the different copies $\Phi _{p}$ of a degenerate field are
considered not as $\left( f,f\right) $ but as $\left( f,j^{p}f\right) $. One
first note that the set $A_{0}$ of the `scalar' fields form a sub-algebra of
the OPA. By `scalar' here we mean not only the spinless fields but also the
single fields under the action of the symmetry $Z_{N}$. A more important
remark is the associativity of (\ref{Fusion2}) which is also a consequence
of the isomorphism with the structure of the chiral fusion rules.

The conservation of the $Z_{N}$ charge $p/N$ in non-chiral fusion rules
leads us to think that associating a simple-current field interpretation
would provide our construction with a less heuristic argument. In the
integer invariant cases, the simple current $\left( j\right) $ is present as
a chiral part of some physical fields, namely: 
\begin{equation}
\,\,\mathbf{J}_{q}^{p}\mathbf{=}\left( j^{q},j^{p+q}\right) \in A_{p}.
\label{Pfields}
\end{equation}
These fields $\mathbf{J}_{q}^{p}$ are indexed by two integers $q$ and $p$.
Indeed, the first index $q$ is present in the two sectors as a power of $j$
and the second index, the \textit{distance} $p,$ is present also as a power
of $j$ but only \ in one sector and so represents the non-diagonal
anisotropy in the coupling between the two sectors. These fields (\ref
{Pfields}) form a sub-algebra of the OPA: 
\begin{equation}
\mathbf{J}_{q_{1}}^{p_{1}}\ast \mathbf{J}_{q_{2}}^{p_{2}}=\mathbf{J}%
_{q_{3}}^{p_{3}};\,q_{3}=q_{1}+q_{2}%
%TCIMACRO{
%\TeXButton{pmod}{\bmod%
%}}%
%BeginExpansion
\bmod%
%
%EndExpansion
N,p_{3}=p_{1}+p_{2}%
%TCIMACRO{
%\TeXButton{pmod}{\bmod%
%}}%
%BeginExpansion
\bmod%
%
%EndExpansion
N,  \label{SF}
\end{equation}
and their action on a field $\left( i,j^{p}i\right) $ $\in A_{p}$ is
determined by our algorithm: 
\begin{equation}
\mathbf{J}_{q_{1}}^{p_{1}}\ast \left( i,j^{p}i\right) =\left(
k,j^{p+p_{1}}k\right) \in A_{\left[ p+p_{1}%
%TCIMACRO{
%\TeXButton{pmod}{\bmod%
%}}%
%BeginExpansion
\bmod%
%
%EndExpansion
N\right] },\,\,\,\,\,k=j^{q_{1}}i.  \label{2}
\end{equation}
Thus, one sees that the physical fields $\,\mathbf{J}_{q}^{p}$ in (\ref
{Pfields}) are the \textit{simple-current fields }of the non-chiral fusion
rules. From their structure, it is easily seen that these simple fields are
divided into two categories. The first one is that of the scalar fields $p=0$%
. They form a sub-algebra of the current-field algebra (\ref{SF}) and are
generated under fusion operation from $\left( j,j\right) $. The second
category is characterized by a non-vanishing distance $p$ and hence is
composed by the set of the non-vanishing spin fields. As a representative
element of this set, one has the holomorphic (antiholomorphic) field $\left(
1,j\right) $ ($\left( j,1\right) $). From (\ref{SF}), it is clear that the
non-vanishing simple spin fields are all generated from $\left( 1,j\right) $
and $\left( j,j\right) $.

As for the chiral fusion rules, the presence of simple fields implies the
conservation of a charge in the non-chiral fusion rules. Indeed, the action
of the simple scalar field $\left( j,j\right) $ (\ref{2}) on a given field
affects the two sectors on equal footing so that the general structure of
this field does not change. Hence, the simple field $\left( j,j\right) $ can
be seen as playing the same role as the simple current $\left( j\right) $
and so it expresses the conservation of the monodromy charge in the
non-chiral fusion rules. This fact comes actually from imposing the
naturality statement in the construction of the non-chiral fusion rules.
More important is the holomorphic simple field $\left( 1,j\right) $ since
its action (\ref{2}) affects only the right sector and hence allows the
establishment of a correspondence between fields in different sets $A_{p}$: 
\begin{equation}
\left( 1,j\right) \ast A_{p}=A_{p+1}.  \label{grad1}
\end{equation}
The holomorphic simple field $\left( 1,j\right) $ plays the role of the
generator of the distance $p$. In terms of \ orbifold partition functions 
\cite{Simple1}, eq. (\ref{grad1}) means that fields in ``twisted sectors'' $%
\left\{ A_{p\neq 0}\right\} $ are generated gradually by the non-chiral
fusion of the ``twist field'' $\left( 1,j\right) $ on \ fields of the
``untwisted sector'' $A_{0}$. From this, we can deduce an important fact
about the behavior of the degenerate fields. Indeed, by denoting each
component of a degenerate field by $\left\{ \Phi _{p}\right\} _{p=0}^{p=N-1}$
as an element of a set $\left\{ A_{p}\right\} _{p=0}^{p=N-1}$ , we can set
from (\ref{grad1}) that: 
\begin{equation}
\left( 1,j\right) \ast \Phi _{p}=\Phi _{p+1}.  \label{3}
\end{equation}
Eq. (\ref{3}) is nothing but a reinterpretation of the action of the $Z_{N}$
cyclic symmetry (\ref{Copies}).

It is now possible to get the general structure of the non-chiral fusion
rules, by establishing their $Z_{N}$-grading structure. First we start
from the general form of the non-chiral fusion rules of the set of `scalar'
fields $A_{0}$: 
\begin{equation}
A_{0}\ast A_{0}=\bigoplus_{i=0}^{N-1}B_{0}^{i}\,A_{i},  \label{A0}
\end{equation}
where the coefficient $B_{0}^{i}$ is non-vanishing if at least one field
from the set $A_{i}$ appears in the non-chiral fusion of two given scalar
fields. Since the scalars are self-conjugate, we must have at least $%
B_{0}^{0}\neq 0$. The general structure of the non-chiral fusion rules can
be deduced by using (\ref{grad1}) as a $Z_{N}$-gradation of the scalar rules
(\ref{A0}): 
\begin{equation}
A_{p}\ast A_{q}=\bigoplus_{i=0}^{N-1}B_{0}^{i}\,A_{i+k},\,\,\,\,k=p+q%
%TCIMACRO{
%\TeXButton{pmod}{\bmod%
%}}%
%BeginExpansion
\bmod%
%
%EndExpansion
N.  \label{Ap}
\end{equation}
It is important at this point to recall from eq. (\ref{3}) that each
component $\Phi _{p}$ of a degenerate field behaves like a field of $A_{p}$
in (\ref{Ap}). In order to complete the determination of the non-chiral
fusion rules, it remains to establish the `scalar' rules (\ref{A0}). It is
important in this regard to recall the nature of the set of `scalar' fields $%
A_{0}$: it represents the ``untwisted sector'' in terms of orbifolding
partition functions. This sector, having an integer charge, forms a
sub-algebra of the original diagonal theory from which the orbifold
procedure has been performed, see \cite{Simple1}. Therefore, it is natural
to set that the scalar rules (\ref{A0}) must have the following `associative' 
structure: 
\begin{equation*}
A_{0}\ast A_{0}=A_{0},
\end{equation*}
and so, finally, the non-chiral fusion rules (\ref{Ap}) will have the
desired $Z_{N}$-grading structure (\ref{Fusion2}).

In conclusion, we have developed in this work a method by which we construct
non-chiral fusion rules as a formal structure describing the operator
content of the OPA of a CFT. We are particularly interested in the
non-diagonal rational conformal models obtained from a simple-current
construction of modular invariant partition functions with cyclic center $%
Z_{N}$. The starting point of this construction is the naturality statement
from which we have imposed the consistency of the non-chiral fusion rules
with the usual fusion rules in the two chiral sectors. The results of these
fusion operations are then combined properly by considering the closure of
the OPA and so the consistency with the operator content which was taken to
be induced by simple currents. In a third step, we have used the simple
current structure of the models to establish the conservation of a charge
termed as \textit{distance charge.} This charge is related to the anisotropy
(the distance) in the coupling between the two sectors via the simple
current $j$. If this symmetry is trivially satisfied in the
automorphism-invariant cases as expected from \cite{Moore}, the situation is
less trivial in the integer-invariant cases and a more careful analysis is
required. For these last cases, the analysis shows the prominent role played
by the holomorphic fields $\left( 1,j^{p}\right) $ as simple current fields
of the distance charge \ symmetry. This important fact reminds us of the
Martinec conjuncture \cite{Marti} which states that the original\ \textit{%
`chiral algebra'} symmetry is not sufficient in order to have a rational
theory, and the symmetry of the full `\textit{set of holomorphic fields'} is
needed in order to accomplish a description of it. As a result, it is
checked \ that each field $\left( i,j^{p}i\right) $ ($Q\left( i\right) =0$)
of the \textit{integer-invariant part} of the operator content will have a
distance charge equal to: 
\begin{equation}
\frac{p}{N}+\alpha \frac{N_{i}}{N}\left( 1-\delta _{N_{i}\,\,N}\right) \,\,%
%TCIMACRO{
%\TeXButton{pmod}{\bmod%
%}}%
%BeginExpansion
\bmod%
%
%EndExpansion
1,\,\,\,\alpha =0\ldots \frac{N}{N_{i}}-1,\text{ \ }p=0\ldots N_{i}-1.
\label{Resu}
\end{equation}
It is important to point out that the second contribution in the distance
charge (\ref{Resu}), present for orbits with length $N_{i}$ $\neq N$, allows
the differentiation between the behaviors of the $N/N_{i}$ copies of a
degenerate primary field $\left( i,j^{p}i\right) $ in the OPA. As a final
result of our construction and by structuring the fields following their
distance charge in sets $A_{p}$, the non-chiral fusion rules have then a $%
Z_{N}$-grading structure.

As a final comment, we recall that the main goal of our construction of the
non-chiral fusion rules is one important step towards the ultimate
determination of the structure constants of the OPA, following our early
work \cite{Amar} for the minimal models. It is therefore of great importance
to be able to find the structure of the non-chiral fusion rules for the more
general cases of non-cyclic center $Z_{p_{1}}\cdot Z_{p_{2}}\cdots Z_{p_{n}}$%
. This is currently under investigation.

\vspace{.2in}

\noindent \textbf{Acknowledgments}

\medskip

\noindent We would like to thank J. Fuchs for his interest in this work and
his constructive correspondence. One of us (A.R.) would like to thank A. Abada 
for all his help and encouragements.

\end{document}